# A Patient-Centric Dataset of Images and Metadata for Identifying Melanomas Using Clinical Context

Veronica Rotemberg[1*], Nicholas Kurtansky[1*], Brigid Betz-Stablein[2], Liam Caffery [2], Emmanouil Chousakos[1,3], Noel Codella[4], Marc Combalia[5], Stephen Dusza[1], Pascale Guitera[6], David Gutman[7], Allan Halpern[1], Harald Kittler[8], Kivanc Kose[1], Steve Langer[9], Konstantinos Lioprys[3], Josep Malvehy[5], Shenara Musthaq[1,10], Jabpani Nanda[1,11], Ofer Reiter[1,12], George Shih[13], Alexander Stratigos[3], Philipp Tschandl[8], Jochen Weber[1], and H. Peter Soyer[2]

* These authors contributed equally to this work
1: Dermatology Service, Department of Medicine, Memorial Sloan Kettering Cancer Center, New York, NY, USA
2: The University of Queensland Diamantina Institute, The University of Queensland, Dermatology Research Centre, Brisbane, AUS
3: First Department of Pathology, University of Athens Medical School, Athens, Greece
4: IBM Research, T. J. Watson Research Center, Yorktown Heights, NY, USA
5: Melanoma Unit, Dermatology Department, Hospital Clínic Barcelona, Universitat de Barcelona, IDIBAPS, Barcelona, Spain
6: Melanoma Institute Australia and Sydney Melanoma Diagnostic Center, Sydney, AUS
7: Emory University School of Medicine, Department of Biomedical Informatics, Atlanta, GA
8: Medical University of Vienna, Department of Dermatology, Vienna, Austria
9: Division of Radiology Informatics, Department of Radiology, Mayo Clinic, Rochester, MN
10: SUNY Downstate Medical School, New York, NY
11: Stony brook Medical School, Stony Brook, NY
12: Rabin Medical Center, Tel Aviv, Israel
13: Department of Radiology at Weill Cornell Medical College

**Corresponding author: Veronica Rotemberg (rotembev@mskcc.org)**

**Figure: 3/3**

**Table: 2/10**

## Abstract

Prior skin image datasets have not addressed patient-level information obtained from multiple skin lesions from the same patient. Though artificial intelligence classification algorithms have achieved expert-level performance in controlled studies examining single images, in practice dermatologists base their judgment holistically from multiple lesions on the same patient. The 2020 SIIM-ISIC Melanoma Classification challenge dataset described herein was constructed to address this discrepancy between prior challenges and clinical practice, providing for each image in the dataset an identifier allowing lesions from the same patient to be mapped to one another. This patient-level contextual information is frequently used by clinicians to diagnose melanoma and is especially useful in ruling out false positives in patients with many atypical nevi. The dataset represents 2,056 patients from three continents with an average of 16 lesions per patient, consisting of 33,126 dermoscopic images and 584 histopathologically confirmed melanomas compared with benign melanoma mimickers.

| | |
|---|---|
| **Design Type(s)** | - database creation objective<br>- data integration objective |
| **Measurement(s)** | - skin lesions |
| **Technology Type(s)** | - digital curation |
| **Factor Type(s)** | - patient identification code<br>- sex<br>- approximate age<br>- general anatomic site<br>- lesion diagnosis |
| **Sample Characteristic(s)** | - Homo sapiens<br>- skin of body |

# Background & Summary

Artificial intelligence (AI) use in medical imaging is rapidly progressing and has the potential to reduce melanoma-associated mortality, morbidity, and healthcare costs by improving access to expertise, diagnostic accuracy, and screening efficiency [1-3]. Here we present a dermatology image dataset that includes patient- and lesion-related clinical context, which can be used in studies to examine whether this additional information further improves recognition performance.

Recent studies have demonstrated the ability of AI algorithms to match, if not outperform, clinicians in the diagnosis of individual skin lesion images in controlled reader studies. Algorithms derived from the 2018 ISIC Grand Challenge have been shown to outperform over 500 clinical readers and experts in such a reader study [1]. However, the reader study did not accurately reflect clinical scenarios where clinicians have access to examine all lesions on a patient.

Clinicians frequently assess skin lesions for biopsy by assessing them in context with the rest of the lesions on a given patient's body, taking into consideration the individual "biologic skin ecosystem". As demonstrated in Figure 1, a lesion with malignancy-predictive features among many similar lesions is thought not to be as dangerous as an odd lesion on a patient whose other lesions are more benign looking. The latter is known in dermatology as the "ugly duckling sign" and is frequently used to diagnose melanoma, especially in patients with multiple melanocytic lesions [4, 5]. Until now, the ugly duckling concept has not been explored with machine learning due to the lack of large datasets with multiple labeled images per patient. Here, we present the first dataset of melanoma and comparative lesions from the same patient to support new machine learning challenges. This dataset is composed of 33126 images collected from 2056 patients at multiple centers around the world such as Memorial Sloan Kettering Cancer Center, New York; the Melanoma Institute Australia and the Melanoma Diagnosis Centre, Sydney; the University of Queensland, Brisbane; the Medical University of Vienna, Vienna; and Hospital

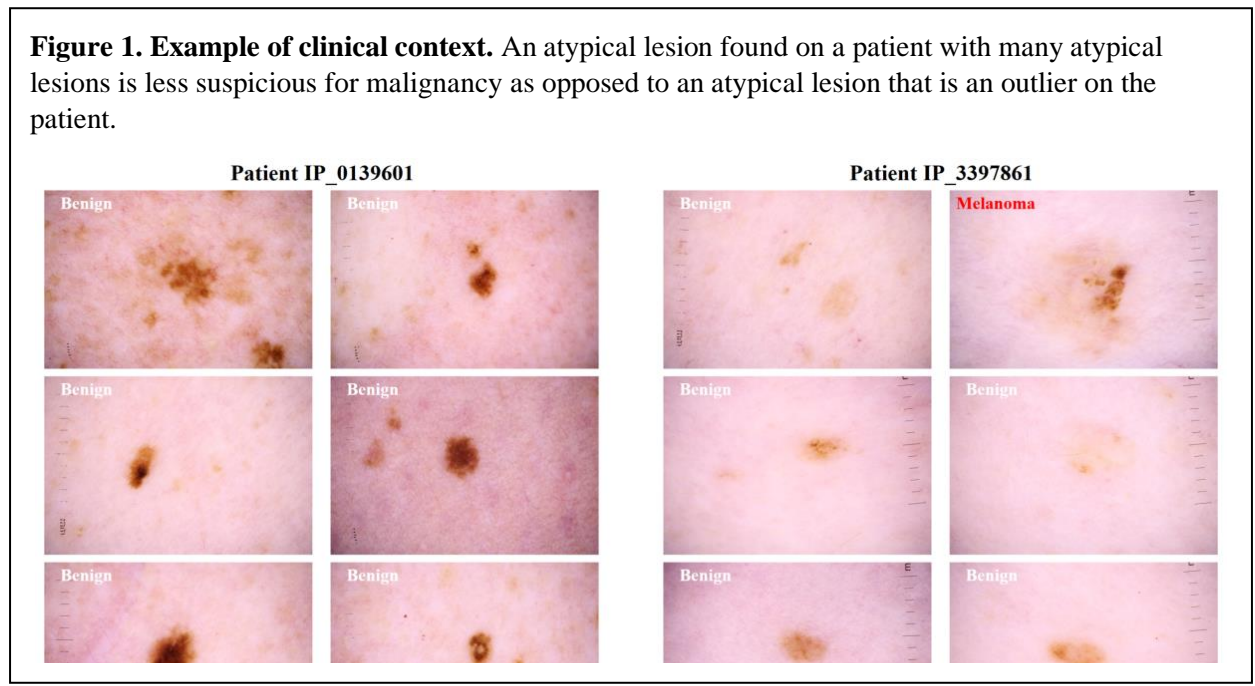

**Figure 1. Example of clinical context.** An atypical lesion found on a patient with many atypical lesions is less suspicious for malignancy as opposed to an atypical lesion that is an outlier on the patient.

Clínic de Barcelona, Barcelona. In this article, we present the methods by which we created this multicenter dataset with clinical contextual information.

## Methods

*General*: We queried clinical imaging databases across the six centers to generate a multicenter imaging dataset. Images consisted of 12,743,090 pixels on average but ranged from 307,200 to 24,000,000. Among patients with dermoscopy imaging from 1998 to 2019, those with multiple skin lesions were identified. Histopathology reports corresponding to internal biopsied lesions were reviewed for diagnosis labelling. Non-biopsied lesions that were monitored for at least six months were considered benign without further granularity [6]. Patients with appropriate qualifying diagnoses: melanoma or benign lesions that could be considered melanoma mimickers including nevi, atypical melanocytic proliferation, café-au-lait macule, lentigo NOS, lentigo simplex, solar lentigo, lichenoid keratosis, and seborrheic keratosis were included [7-9]. Lesions satisfying the described criteria were represented in the dataset with a single dermoscopic image [8, 10, 11]. These include images captured with or without polarized light using a contact or noncontact dermatoscope. When multiple image types were available, the selected image was either the one of highest resolution or if multiple images at the same resolution were available, one was chosen randomly. Images containing any potentially identifying features, such as jewelry or tattoos, or from patients without at least three qualifying images were excluded during quality assurance review.

In order to test algorithm generalizability, a subset of images from six sites (five geographic locations) were allocated for the training dataset of the 2020 ISIC Grand Challenge [12].

*Quality Assurance*: A software annotation tool, called '*Tagger,*' was developed internally to review diagnostic labeling of grouped images [13]. Using this tool, dermoscopy expert reviewers (EC, OR) were presented sets of 30 images with a shared diagnosis in order to identify the ones with erroneous labeling. Reviewers invested 22 hours over three weeks of quality assurance in *'Tagger'* and spent an average of 4 seconds per set when flagging a single image, and 11 seconds per set when flagging several images. Out of all images reviewed in *Tagger*, 2.7% were removed, out of concern for erroneous labels.

*Memorial Sloan Kettering Cancer Center*: The MSK Dermatology Service is a high-risk clinic that relies heavily on imaging for high risk individuals with or without a history of melanoma [14]. Images were acquired using a dermoscopic attachment to either a digital single reflex lens (SLR) camera or to a smartphone. Each lesion was imaged with polarized and/or nonpolarized dermoscopy. For each lesion, 3-5 images are collected during each patient visit and stored in a specialized image database called Vectra™ (Canfield Scientific Inc., Parsippany, NJ, USA).

Images were extracted after searching the database for patients with multiple lesions imaged and who had biopsy confirmed melanoma from 2015-2019. The clearest image per time point was selected by medical student research fellows using a selection tool designed uniquely for the task (SM, JN, OR, EC) [15].

*Hospital Clínic Barcelona*: The Department of Dermatology of the Hospital Clínic of Barcelona is a tertiary referral center for melanoma patients, includes a high-risk melanoma patient clinic. The dermatology department is equipped with the digital dermatoscopy system MoleMax™ HD (Derma Medical Systems, Vienna, Austria) and corresponding image database to store the collected images. Each lesion was photographed using polarized dermoscopy.

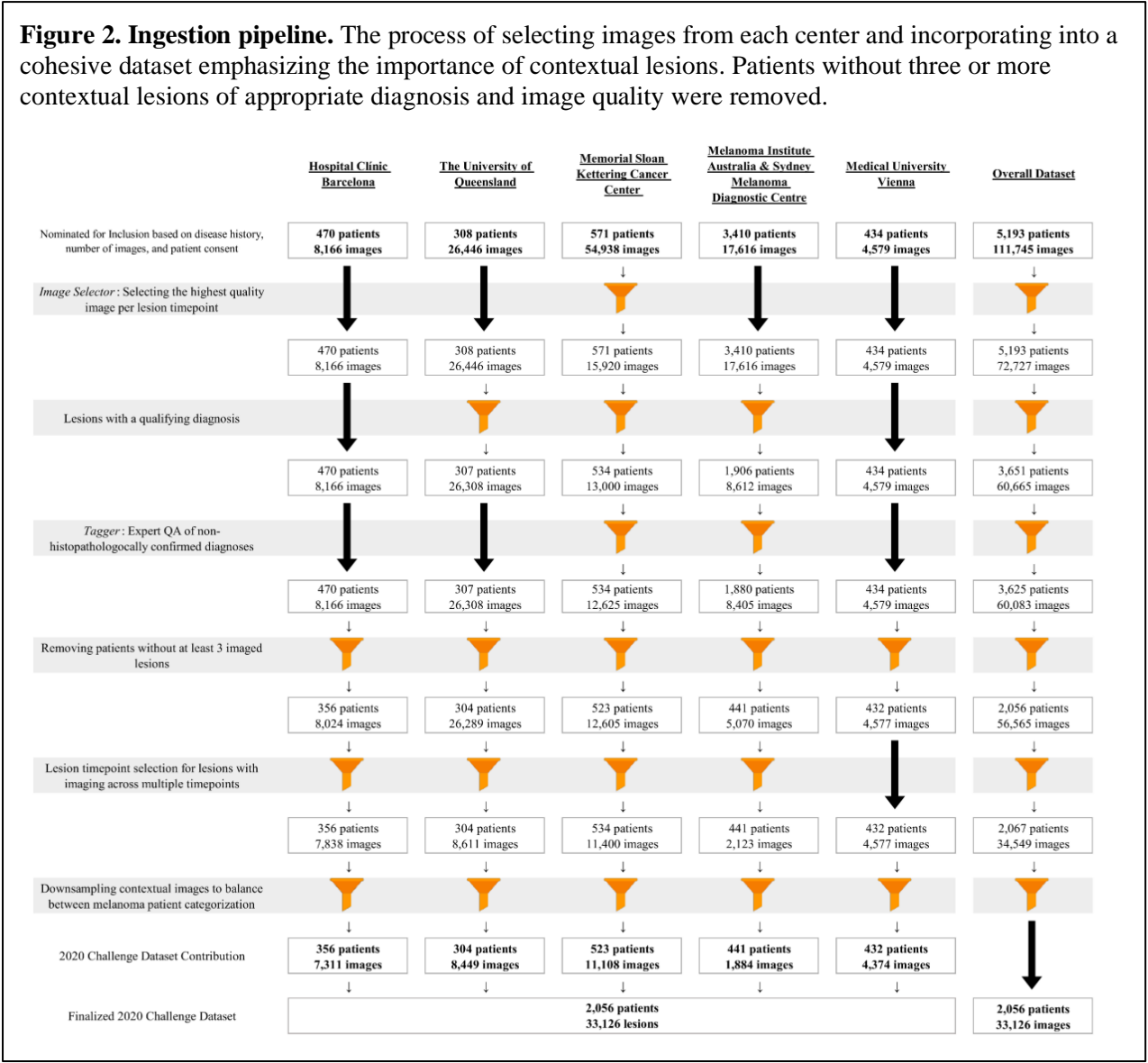

**Figure 2. Ingestion pipeline.** The process of selecting images from each center and incorporating into a cohesive dataset emphasizing the importance of contextual lesions. Patients without three or more contextual lesions of appropriate diagnosis and image quality were removed.

Candidate images were extracted after searching the database for benign lesions with >1.5 years of digital dermoscopy follow-up and excised lesions with a histopathology report. The images were examined by three expert dermatologists at the clinic for image quality assurance and label accuracy. From a series of multiple sequential images of the same nevus, we extracted the median timepoint.

*The University of Queensland*: The Clinical Research Facility of the Translational Research Institute in Brisbane, Queensland, Australia is the clinical trial site following both general population and high-risk individuals participating in studies carried out by the Dermatology Research Center of The University of Queensland Diamantina Institute. Contributed images came from three prospective longitudinal studies. The first study, "Changing Naevi Study", consisted of two groups of participants; advanced stage (III – IV) melanoma patients undergoing treatment with immunotherapy and/or targeted therapy and people at high risk of developing melanoma due to personal or family history but were not undergoing treatment at time of enrollment. The second study, "Mind Your Moles", consisted of participants from a general population cohort recruited from the Brisbane Electoral Role [16]. All nevi >5 mm were imaged in these two studies, as well as any lesions of interest/concern to the participant or clinician. The third study, "Evaluation of the Efficacy of 3D Total-Body Photography With Sequential Digital Dermoscopy in a High-Risk Melanoma Cohort", consisted of participants at high risk of melanoma [17], half of which underwent imaging intervention [18]. Lesions of interest to the participant or clinician were imaged dermoscopically.

All images used for the studies were extracted from the Vectra™ image database (Canfield Scientific Inc., Parsippany, NJ, USA).

*Medical University Vienna*: The Early Recognition Unit of the Department of Dermatology of Medical University of Vienna is a tertiary referral center for high-risk patients. It offers total digital dermatoscopic follow-up to patients with multiple nevi [19]. Most patients in the program are of European descendance with fair skin types (usually skin type 1-3) and have a high number of nevi and a personal or family history of melanoma.

We extracted polarized dermoscopic images from 2015-2019 which were stored in the MoleMax HD System (Derma Medical Systems, Vienna, Austria). We searched the database of this system for patients with at least 3 dermoscopic images by filtering SQL-tables with a proprietary tool provided by the manufacturer. From these patients we selected all benign melanocytic lesions with > 1 year follow-up and all lesions that were excised. Histopathology reports were matched manually to all excised lesions. Non-melanocytic lesions, duplicate images, images captured before 2015 with older systems, low-quality images, and images that depicted only parts of the lesion were excluded. Furthermore, we excluded images of lesions that were already included in the 2018 or 2019 ISIC challenges.

*Melanoma Institute Australia and the Sydney Melanoma Diagnosis Centre*: Both services are high-risk, tertiary referral dermatology clinics that rely heavily on imaging of individuals with or without a history of melanoma [14]. Lesions imaged for short term monitoring are selected at the discretion of the clinician or which are of concern to the patient. Additionally, all lesions are imaged prior to surgical removal. Images are acquired using a dermoscopic attachment to either a digital single reflex lens (SLR) camera or to a smartphone and stored in DermEngine™ (Metaoptima, Vancouver, British Columbia, Canada). Histopathology reports were reviewed and lesions followed for six months or more without malignant changes were considered benign. All images were manually reviewed to assure de-identification and image detail quality after database extraction.

*Dataset Compilation*: The quality assurance and collection steps we performed for curating the images from various sources are detailed in Figure **2**2.

*Lesion Timepoints*: Each lesion in the dataset is represented by a single image. The image of non-biopsied benign lesions with imaging at multiple time points were selected to minimize the difference in patient imaging date variability and date range between patients with and without an imaged melanoma. This was performed to reduce potential bias in image lighting, camera type, or other factors between the benign and melanoma patient class.

*Lesion Context Images*: Due to the retrospective nature of image acquisition and potential surveillance bias in different patient populations, the number of lesions per patient was not distributed identically between the class of patients with a melanoma image and the class without a melanoma image. Because the lesions in this dataset do not represent all lesions that exist on this set of patients, it is possible the imbalance is related to selection bias of imaged lesions. Lesions in both classes were subsampled through patient matching, which led to a loss of 4.1% of images. Ultimately, 50% of the lesions have more than 10 contextual lesions. The matched number of images per patient ID before and after subsampling is shown in Figure 3.

*Duplicates*: Due to a clerical error during the data ingestion process to the ISIC Archive, 425 pixelwise identical duplicate images were ingested and included in the dataset. The 2020 SIIM-ISIC Melanoma Classification competition page on Kaggle lists the redundant cases and deidentified patient labels (https://www.kaggle.com/c/siim-isic-melanoma-classification/discussion/161943). This information is also available upon request to the authors.

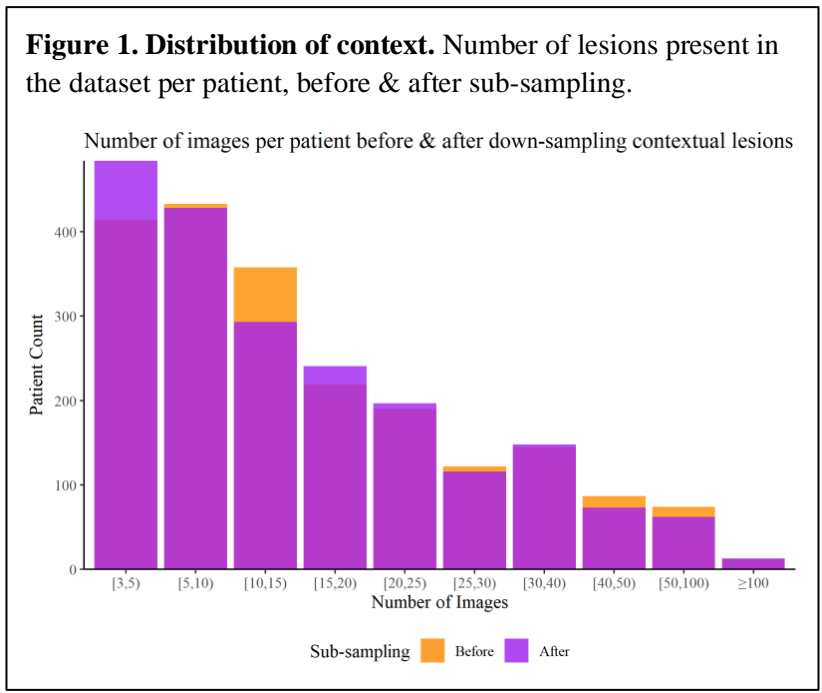

**Figure 1. Distribution of context.** Number of lesions present in the dataset per patient, before & after sub-sampling.

## Data Records

The dataset will be available for download through the Kaggle platform as part of a live competition through August 20, 2020. It is governed by the ISIC Terms of Use and licensed as a CC-BY-NC, and will permanently be accessible to the public after the 2020 challenge is complete through the ISIC Archive at this DOI: https://doi.org/10.34970/2020-ds01. Currently no modifications have been made to the dataset, however, any metadata or image modifications will be noted at that DOI landing page.

Metadata for each image included approximate patient age at time of image capture, biological sex, general anatomic site of the lesion, anonymized patient identification number, benign/malignant category, and the specific diagnosis if one was available based on an acceptable ground truth confirmation method. A summary of the characteristics of the dataset at patient- and lesion-level is shown in Table 1.

*Dataset format:* The dataset is available in two formats.

The first is the file format described in Part 10 of Digital Imaging and Communication in Medicine (DICOM) standard [20, 21]. The DICOM standard is a comprehensive, international medical image standard that was originally developed for radiology, where it has become ubiquitous as the core standard. It has since been adopted by many other medical imaging specialties including ophthalmology, dentistry, cardiology, nuclear medicine, oncology, pathology, surgical specialties who perform image-guided surgery (e.g., neurosurgery, ENT, orthopedics), and specialties that acquire endoscopic or laparoscopic imaging [22]. The DICOM file format is an amalgamation of the metadata and pixel data in a single file. The pixel data is encoded in Joint Photographic Expert Group (JPEG) format.

The second format is where the images are either in JPEG or Tagged Image File (TIF) format and the metadata is included in a linked comma-separated values (CSV) file.

## Technical Validation

The ground truth labels for all malignant lesions in the dataset were confirmed via retrospective review of histopathology reports, and diagnosis plausibility was visually confirmed by visual confirmation of a dermoscopy expert. Histopathology reports were double checked if the label was suspicious. Melanoma in situ and invasive melanoma were both coded as melanoma. All other qualifying images were coded as benign, including those diagnosed as severely dysplastic nevi [23, 24].

Non-biopsied lesions with expert consensus agreement and lesions followed for six months or more without malignant changes were labelled benign without a more specific diagnosis by most contributors. Dermatofibromas, seborrheic keratosis, or vascular lesions were not monitored, as that would not reflect clinical practice, but labels were verified visually by an expert in dermoscopy. Images of lesions were attributed to patients based on the clinical imaging database identification codes which are stored at the time of capture during each clinical photography session.

## Usage Notes

This dataset is the first dataset that mimics clinical practice by labeling images (mean = 16, median = 12, standard deviation = 16) from the same patient as such and allowing algorithms to assess multiple images from the same patient for malignancy. It addresses a particularly challenging area of clinical practice, those patients with multiple atypical nevi suspicious for malignancy. The dataset is designed to improve translational potential of algorithms, especially to help clinicians without access to tertiary referral centers assess high risk patients with multiple atypical nevi. Additionally, algorithms developed using this dataset may be better candidates for incorporating into dermatology imaging systems, as they can evaluate all images for a given patient in context, and perhaps even be used during clinic visits in which multiple lesions are imaged. Given the translational potential of algorithms developed using this dataset, we hope that generating a public, well annotated dataset that mimics clinical practice will lead to prospective studies of promising automated approaches for diagnosing melanoma.

Various forms of dermoscopy imaging are included in the dataset: contact non-polarized light, contact polarized light, and non-contact polarized light. Deeper skin structures are more often visible under polarized light than non-polarized light, even without direct skin contact with the interface or the use of a liquid interface [25]. Various colors, structures, and patterns are more pronounced, or accessible, under specific forms of dermoscopy [26]. Imaging modalities are not equivalent in identifying certain morphologies but complement one another in a holistic clinical assessment. A limitation to this dataset is that each lesion is represented by a single image and type of dermoscopy, which may not reflect the full spectrum of information that would be used by a clinician [22].

Generalization of AI-assisted skin lesion classification to broad clinical use depends on the demographic agreement of the training dataset to the clinical population. Due to low population prevalence and challenges with access to care in different populations, the images gathered for large datasets such as this for AI classification have a strong tendency to under-represent darker skin types. This may lead to either overdiagnosis or underdiagnosis of melanomas in darker skin types, both of which would have significant clinical implications and will require prospective study. The ISIC Archive is actively pursuing methods by which to increase the diversity of images obtained, but at this point caution should be used when attempting to generalize algorithms trained on images from specialized referral centers (such as the dataset described herein) to the global population at large. The dataset is also enriched for melanoma in general and does not represent true incidence of melanoma.

**Table 1. Summary of combined dataset.** Patient- and lesion-level characteristics of the dataset.

|  | Patient Class | | |
|---|---|---|---|
|  | ≥ 1 Melanoma | 0 Melanoma | Total |
| Patients (n) | 428 | 1628 | 2056 |
| Male | 260 | 817 | 1077 |
| Female | 168 | 809 | 977 |
| Unknown | 0 | 2 | 2 |
| Average Age (years) | 57.7 | 49.6 | 51.3 |
| Lesions (n) | 6927 | 26199 | 33126 |
| Benign (Not Biopsied) | 5862 | 25256 | 31118 |
| Benign (Biopsied) | 481 | 943 | 1424 |
| Melanoma | 584 | 0 | 584 |
| Lesions per Patient |  |  |  |
| Mean | 16.2 | 16.1 | 16.1 |
| Q1 | 5 | 5 | 5 |
| Median | 12 | 12 | 12 |
| Q3 | 22 | 22 | 22 |
| Timepoints per Patient |  |  |  |
| Mean | 5.9 | 3.4 | 3.9 |
| Q1 | 3 | 1 | 1 |
| Median | 4 | 2 | 3 |
| Q3 | 7 | 4 | 5 |
|  | Lesion Diagnosis | | |
|  | Melanoma | Benign | Total |
| Lesions per Patient Class (n) |  |  |  |
| ≥ 1 Melanoma | 584 | 6343 | 6927 |
| 0 Melanoma | 0 | 26199 | 26199 |
| Lesions per Anatomic Site (n) |  |  |  |
| Head/Neck | 74 | 1781 | 1855 |
| Torso | 257 | 16588 | 16845 |
| Upper Extremity | 111 | 4872 | 4983 |
| Lower Extremity | 124 | 8293 | 8417 |
| Palms/Soles | 5 | 370 | 375 |
| Oral Genital | 4 | 120 | 124 |
| Unknown | 9 | 518 | 527 |

## Code Availability

Custom generated code for the described methods is available at https://github.com/ISIC-Research/2020-Challenge-Curation.

## Acknowledgements

The dataset provided by The University of Queensland in Brisbane was funded by the National Health and Medical Research Council (NHMRC) – Centre of Research Excellence Scheme (APP 1099021). HPS holds an NHMRC MRFF Next Generation Clinical Researchers Program Practitioner Fellowship (APP1137127). Other funding sources include the Melanoma Research Alliance Young Investigator Award 614197 and NIH Core Grant (P30 CA008748).

## Competing Interests

HPS is a shareholder of MoleMap NZ Limited and e-derm consult GmbH, and undertakes regular teledermatological reporting for both companies. HPS is a Medical Consultant for Canfield Scientific Inc., Revenio Research Oy and also a Medical Advisor for First Derm. AH serves as a consultant to Canfield Scientific Inc., and is on the SciBase advisory panel.